\documentclass[preprint,pra,aps,amsmath,amssymb,showpacs,superscriptaddress]{revtex4-2}

\usepackage{graphics}
\usepackage{epsfig}
\usepackage{hyperref}
\usepackage{amsfonts}
\usepackage{amssymb}
\usepackage{amsmath}
\usepackage{color}
\usepackage[usenames,dvipsnames]{xcolor}

\usepackage{graphicx}
\usepackage{dcolumn}
\usepackage{bm}

\begin{document}
\title{Quantum degeneracy in mesoscopic matter: Casimir effect and Bose-Einstein condensation.}
\author{I.~Todoshchenko}
\author{M.~Kamada} 
\author{J.-P.~Kaikkonen} 
\author{Y.~Liao} 
\author{A.~Savin} 
\author{E.~Kauppinen} 
\author{E.~Sergeicheva} 
\author{P.J.~Hakonen}
\affiliation{Low Temperature Laboratory,
Deptartment of Applied Physics, Aalto University University, 00076~AALTO, Finland}
\email{igor.todoshchenko@aalto.fi}
\date{\today}

\begin{abstract}
The ground-state phonon pressure is an analogue to the famous Casimir pressure of vacuum produced by zero-point photons. The acoustic Casimir forces are, however, many orders of magnitude weaker than the electromagnetic Casimir forces, as the typical speed of sound is 100 000 times smaller than the speed of light. Because of its weakness, zero-point acoustic Casimir pressure was never observed, although the pressure of artificially introduced sound noise on a narrow aperture has been reported. However, the magnitude of Casimir pressure increases as $1/L^3$ with the decrease of the sample size $L$, and reaches picoNewtons in the sub-micron scales. We demonstrate and measure the acoustic Casimir pressure induced by zero-point phonons in solid helium adsorbed on a carbon nanotube. We have also observed Casimir-like "pushing out" thermal phonons with the decreasing temperature or the length. We also show that all thermodynamic quantities are size-dependent, and therefore in the mesoscopic range $L\lesssim\hbar{c}/(k_BT)$ quadruple points are possible on the phase diagram where four different phases coexist. Due to the smallness of solid helium sample, temperature of Bose-Einstein condensation (BEC) of vacancies is relatively high, $10-100$\;mK. This allowed us to experimentally discover the BEC in a system of zero-point vacancies, predicted more than 50 years ago.
\end{abstract}

\maketitle
\section*{Introduction}
Casimir effect is one of the most counter-intuitive phenomenon of various manifestations of quantum mechanics. It was first formulated by Lars Casimir \cite{Casimir1948} as a pressure of vacuum, i.e.~of empty space containing no material particles. According to Casimir, zero-point fluctuations related to electromagnetic modes between two parallel plates lead to attractive force between the plates. Equivalently, zero-point fluctuations for other excitations, e.g.~phonons, can lead to Casimir-like effects.

In general, there exists three different types of Casimir-like effects. First, the ordinary Casimir force which is a quantum mechanical effect due to zero-point photons/phonons. The predictions for electromagnetic Casimir effect were verified in beautiful experiments in 1997 \cite{Casimir1997}. Second, Casimir effect in which thermal excitation of modes plays an important role, and the Casimir force scales with temperature \cite{Milton}. Third, Casimir-like phenomena which are generated by broad band noise that is not able to propagate to sections of interest in the studied geometry. Acoustic Casimir effect has so far been only demonstrated using external noise \cite{Larraza1998}. In fact, this kind acoustic noise Casimir pressure has been found almost 200 years ago by noticing the attraction of two closely placed ships \cite{ships}. However, genuine acoustic analogue of the zero-point Casimir pressure has never been observed prior to our work. Our measurements indicate Casimir pressure due to zero-point motion of quasi one-dimensional phonons in solid $^3$He on a carbon nanotube.

The acoustic zero-point Casimir pressure is many orders of magnitude tinier than the electromagnetic Casimir pressure because of the smallness of speed sound compared to speed of light. However, the Casimir forces rapidly increase  with the decrease of the length of the sample. Fortunately, helium adsorbed on a carbon nanotube (CNT) represents a sample which is small enough for Casimir forces to be significant, and still large enough to be measured.

\section{Experimental}

We have studied mechanical resonances of a suspended carbon nanotube (CNT) with adsorbed sub-monolayer of $^3$He. The CNTs were synthesized in the gas phase with the floating catalyst chemical vapour deposition growth method (FC-CVD) followed by the direct thermophoretic deposition onto prefabricated chips \cite{Wei2019}. Radius of the tube $r_0=0.8$\;nm and length $L=700$\;nm has been measured with electron microscope.  Electro-mechanical scheme of the measurements described in the Appendix. After a detailed electrical characterization of the suspended CNT along with its mechanical resonance properties in vacuum, $^3$He atoms were gradually added to the sample chamber. As a single nanotube can adsorb only very little amount of $^3$He, 1-10 thousands atoms, we used a grafoil ballast \cite{grafoil} with the surface area 10\;m$^2$. Using grafoil, we were able to control the coverage of the adsorbed $^3$He with the accuracy of 0.03\;$\text{nm}^{-2}$. 

\begin{figure}[htb]
\includegraphics[width=0.99\linewidth]{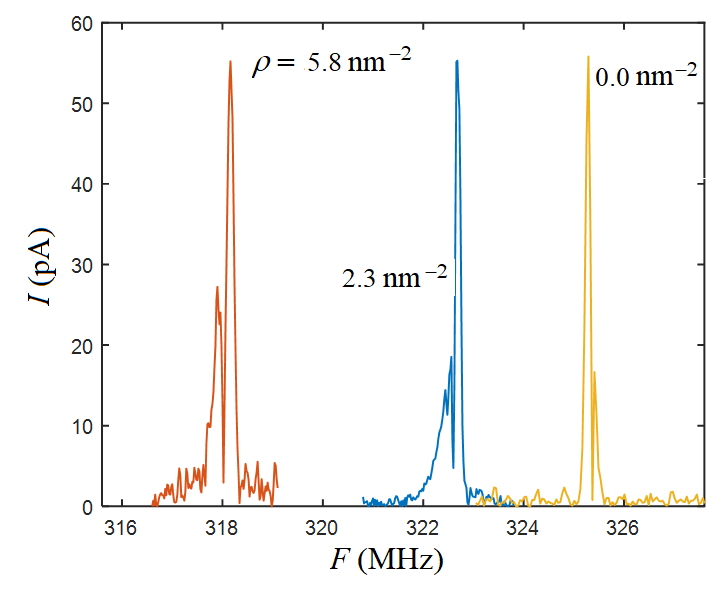}
\caption{\label{fig:resonances} Determination of helium coverage on the suspended carbon nanotube at 0.25\;K. The shift of the resonance is due to increase in the mass of the oscillating nanotube, while the additional stiffness can be ignored because helium is in a fluid phase.}
\end{figure}

After each addition of $^3$He atoms, we waited for 10-20 hours and, subsequently, performed temperature sweeps up and down at a rate of 10\;...\;30\;mK$/$h. Typically, the $T$ sweep at each coverage was repeated at least once, sometimes several times. The coverage of helium $\rho$ was determined by measuring the shift of the resonance frequency when helium was in fluid phase at high enough temperature, see Fig.\;\ref{fig:resonances}.

In constructing our coverage scale
\begin{equation}
\label{eq:coverage_liq}
\frac{\Delta F}{F_0}=-\frac{M_{He}}{2M_C}=-\frac{\rho_{He}}{8\rho_{C}} 
\end{equation} 
we have assumed, that the change in the elastic state (stiffness) of the tube can be neglected in fluid phases at high temperatures, and that the frequency shift is governed by the total mass of helium atoms ${M_{He}}$ on the CNT compared 
with the mass of the carbon lattice ${M_C}$. At high $T$, besides being in the fluid phase, the distance between helium atoms and the substrate is the largest, which minimizes the influence of helium atoms on the elastic properties of the CNT. Upon localization of the atoms when entering solid phases, the free surface energy becomes lowered, which leads to a change in the spring constant of the tube as observed in the experiments as a frequency shift \cite{CNT3HE}.

\section{Spectroscopy of phonons: Casimir effect and the fate of thermodynamics in mesoscopic matter}

Similarly to the pressure of vacuum due to zero-point photons, there is a pressure in condensed matter due to zero-point phonon modes. As in the Casimir effect, this pressure depends on the size of the sample, because the larger the sample, the longer wavelengths are allowed. Owing to the presence of long-wave photons (phonons), which do not fit the smaller volume, the larger volume has a higher pressure compared with the small one. Helium adsorbed on the nanotube is a perfect model system to investigate the size dependence of zero-point pressure, a full analog of the Casimir force. The lack of the phonon zero-point energy in the sample of the length $L$ is the energy of the phonons with the wavelength $\lambda>L$,     $\Delta{E}_0=-m(\hbar/2)\sum_{k_n=\pi/\mathcal{L}}^{\pi/L}{c}k_n=   -m(\hbar{c}/2)\sum_{n=1}^{\mathcal{L}/L}(\pi/\mathcal{L})n= -m\hbar{c}\pi\mathcal{L}/4L^2$, where the long wavelength cutoff $\mathcal{L}\approx10$\;cm is the characteristic size of helium layer outside the nanotube. As helium on nanotube is not a true 1D system but has 20-25 atoms on its cross-sectional circumference, we include a corresponding multiplicity $m$ for each longitudinal phonon mode. The acoustic Casimir effect differs from the traditionally considered electromagnetic Casimir effect because of the existence of the short wavelength cutoff (interatomic distance), and the long wavelength cutoff (size of the sample) for phonons. This results in convergence for the total sum of phonon energies, in contrast to the photon case and, consequently, in a different analytical form of the Casimir energy.

At $T=0$ the Casimir force acting on the tube is found simply as $\mathcal{F}_{\rm C}=\partial\Delta{E}_0/\partial{L}=m\hbar{c}\pi\mathcal{L}/2L^3$. The power $L^{-3}$ instead of the $L^{-2}$ in ordinary 1D Casimir force \cite{Milton} is due to the finite cutoffs in the case of phonons, while there is no cutoffs for photons. 
1D pressure $P$ of helium adsorbed on the tube, which is the force acting on the circumference at the boundary, contributes directly to the tension $\mathcal{F}_0$, $\mathcal{F}=\mathcal{F}_0+P$. Indeed, 1D pressure of adsorbed helium is the line energy density, $P=dE_{He}/dL$, whilst the line tension of the tube is the elastic energy density, $\mathcal{F}_0=dE_{CNT}/dL$. Physically, it can be understood as the force produced on the tube by reflection of individual (quasi-) particles from boundaries. Casimir force reduces the pressure of helium on the tube and, accordingly, the tension, $\mathcal{F}=\mathcal{F}_0+P-\mathcal{F}_{\rm{C}}$. By measuring the resonant frequency $F_0=(1/2L)\sqrt{\mathcal{F}/\mu}$ we can therefore detect the Casimir force. Due to the smallness of the tube, the phonon modes are mostly in the ground state up to temperature $T_{ZP}(L)=\pi\hbar{c}/L\approx20$\;mK for $L=700$\;nm, and up to higher temperatures for smaller solid fractions.

At lower average densities $\rho<\rho_{1/3}$ the solid phase occupies only a part of the tube with the length $L_{1/3}=L(\rho-\rho_L)/(\rho_{1/3}-\rho_L)$, while the rest is liquid with low density $\rho_L$. This configuration is energetically more favorable than the alternation of many small phases "solid-liquid-solid...", the so-called Domain wall solid, suggested in Refs. \onlinecite{HalpinKardar1986,Godfrin1995,Morishita2001} because in the latter case there appear many interfaces between the regions of different phases, which have additional energy due to the surface tension. 

Historically, the term "acoustic Casimir effect" was introduced for the effect of mutual force between two closely placed plates under the sound noise generated in the chamber by a microphone \cite{Holmes1997,Larraza1998}. Depending on the spectrum of the generated noise, the force could be attractive or repulsive. In contrast, in our experiment we deal with a complete analogue of the Casimir pressure: the pressure is due to acoustic vacuum, to which only ground-state phonons with energy $\hbar\omega/2$ contribute. The contribution of thermal phonons in the tube is negligible at $T\lesssim10$\;mK. Pressure of thermal phonons in the surrounding bath $F_{Bath}=\pi{m}k_B^2T^2/12\hbar{c}\simeq10^{-17}$\;N is also negligible. The effect is achieved by reducing the phonon phase space using a small sample, which is exactly along the lines of experiments targeting the original Casimir pressure \cite{Casimir1997}.

\begin{figure}[htb]
\includegraphics[width=0.98\linewidth]{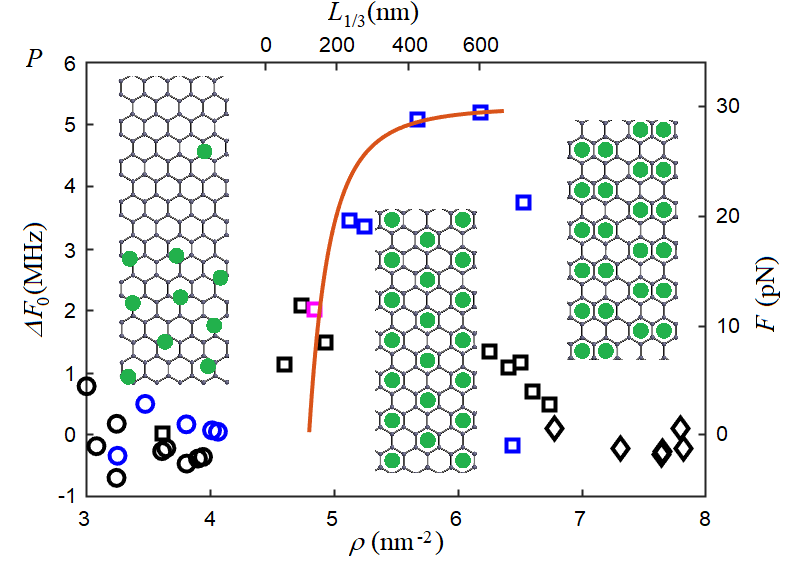}
\caption{\label{fig:phonons0} Resonance frequency shift $\Delta F_0$ due to 1D pressure (force) of zero-point phonons ($T<20$\;mK) as a function of atomic coverage per nm$^2$ (lower axis) and the corresponding length occupied by the 1/3 solid, $L_{1/3}=L(\rho-\rho_L)/(\rho_{1/3}-\rho_L)$ (higher axis). Curve displays the contribution of the Casimir force, $\mathcal{F}_{\rm C}=50\hbar{c}\pi\mathcal{L}/L^3$, see text for details.}
\end{figure}

Fig.\;\ref{fig:phonons0} shows the dependence of the central frequency $F_0$
on the density of helium at the lowest $T=10$\;mK. With the increase of density, the helium layer transforms from liquid below $\rho=4.1$\;nm$^{-2}$ to the so-called soft 2/5 solid, through a mixture of solid 1/3 phase and liquid upto $\rho_{1/3}=6.4$\;nm$^{-2}$. There are two major effects of adsorbed helium on the resonance frequency of the tube: increase due to elastic stiffening and pressure, and decrease due to additional mass. After subtracting the linear trend in frequency due to the increase of helium mass, we are left with contributions of helium-induced stiffness and pressure. In the liquid phase, the pressure is obviously zero as it is in contact with exponentially diluted gas, as shown in the leftmost insert in Fig.\;\ref{fig:phonons0}. Indeed, the length of the 1/3 solid shown on the top axis has a dramatic effect on phonon pressure, which is illustrated by the solid curve $\mathcal{F}_{\rm C}=m\hbar{c}\pi\mathcal{L}/L^3$ using the multiplicity $m=50$ and an offset corresponding to the additional tension produced by solid helium in the limit of large samples, $L\rightarrow\infty$. This offset tension of long samples, at densities close to $\rho\rightarrow\rho_{1/3}=6.4$\;nm$^{-2}$, is set by the attraction of the adsorbed helium atoms to neighboring carbon atoms \cite{CNT3HE}. The opposite limit, $L\rightarrow0$ is not realized physically as it corresponds to a negative effective pressure due to the Casimir force (see above, $P-\mathcal{F}_{\rm{C}}$ becomes $<0$), which is not stable. Liquid phase at $\rho<\rho_L=4.1$\;nm$^{-2}$ is in contact with strongly rarefied gas, and thus has zero pressure. Therefore, when 1/3 solid starts to form on the tube, it has zero pressure at $\rho=\rho_L\simeq4.8$\;nm$^{-2}$ which increases at larger average densities. In the range 4.1\;nm$^{-2}<\rho<4.8$\;nm$^{-2}$, there appears a gap in the phase diagram which can be assigned to the described negative pressure instability. Indeed, during the experiment we could not fill this domain of coverages, while we were able to fill neighboring areas with steps in coverage less than 0.05\;nm$^{-2}$.

The abrupt drop at $\rho=\rho_{1/3}$ indicates a quantum phase transition to the soft 2/5 solid ($\rho_{2/5}=7.6$\;nm$^{-2}$) which does not support phonons \cite{CNT3HE}. 
Mobile defects have very low zero-point energy, of the order of $\hbar^2/(2mL^2)\sim1$\;$\mu$K, with vanishing contribution to pressure at low temperatures. Zero-point pressure of a commensurate solid phase is thus only due to zero-point phonons which are absent in the 2/5 phase whose characteristic pressure is therefore nearly zero. Hence, at $\rho>\rho_{1/3}$ a mixture of 2/5 phase and incommensurate solid is realized. As pressure must be constant over the tube, density of the incommensurate phase tunes by the unique pressure of the 2/5 phase.

At finite temperatures thermal phonons are created, in addition to zero-point phonons. However, the small size of the sample restricts greatly the amount of thermal phonons. In the limit $L\ll\pi\hbar{c}/T$ the free energy $F$, entropy $S$, energy $E$, and pressure $P$ of 1D phonons with multiplicity $m$ of each mode are written as

\begin{eqnarray}
\label{eq:thermo}
F=\frac{mLk_BT}{2\pi{c}}\int_{\pi{c}/L}^{\infty} \ln{[1-e^{-\hbar\omega/k_BT}]}d\omega=        -\frac{m Lk_B^2T^2}{2\pi\hbar c}\int_{\pi\hbar c/Lk_BT}^{\infty} \ln{[1-e^{-x}]}dx= \nonumber \\
-\frac{mLk_B^2T^2}{2\pi\hbar{c}}\int_{\pi\hbar{c}/Lk_BT}^{\infty} \frac{xdx}{e^{x}-1}\simeq                   -\frac{mLk_B^2T^2}{2\pi\hbar{c}}\int_{\pi\hbar{c}/Lk_BT}^{\infty}     xe^{-x}dx=-\frac{mLk_B^2T^2}{2\pi\hbar{c}}              \Big{(}1+\frac{\pi\hbar{c}}{2Lk_BT}\Big{)} e^{-\pi\hbar{c}/Lk_BT} \nonumber \\
S=-\frac{dF}{dT}\Big {|}_L\simeq                   m\Big{(}\frac{Lk_B^2T}{\pi\hbar{c}}+\frac{3k_B}{4}+\frac{\pi\hbar c}{4LT}     \Big{)}e^{-\pi\hbar{c}/Lk_BT} ~~~~~~~~~~~~~~~~~~~~~~~~~~~~\\
E=F+TS\simeq m\Big{(}\frac{Lk_B^2T^2}{2\pi\hbar{c}}+\frac{k_BT}{2}+ \frac{\pi\hbar{c}}{4L}\Big{)}e^{-\pi\hbar{c}/Lk_BT}~~~~~~~~~~~~~~~~~~~~~~~~~ \nonumber \\
P=-\frac{dF}{dL}\Big {|}_T\simeq  m\Big{(}\frac{k_B^2T^2}                               {2\pi\hbar{c}}+\frac{k_BT}{2L}+ \frac{\pi\hbar{c}}{4L^2}\Big{)}e^{-\pi\hbar{c}/Lk_BT}.~~~~~~~~~~~~~~~~~~~~~~~~ \nonumber
\end{eqnarray}

\noindent
As expected, at $Lk_BT\ll\hbar{c}$ all thermodynamic potentials are exponentially small. Practically, this means that there are no phonons except for zero-point ones, and energy and pressure in the sample are temperature-independent.

\begin{figure}[htb]
\includegraphics[width=0.98\linewidth]{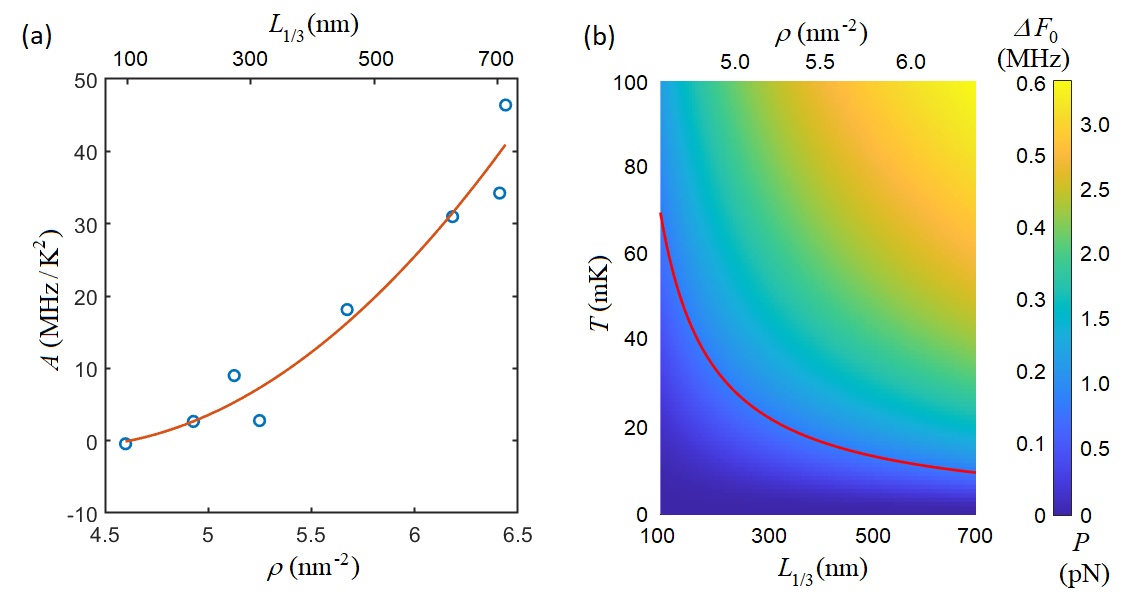}
\caption{\label{fig:phonons1} (a) Pre-factor $A$ of the fit of the resonant frequency shift $\Delta{F_0}=AT^2$ as a function of coverage/length of 1/3 phase. Red curve is a quadratic fit. (b) Smoothed data on pressure of thermal phonons (in terms of frequency shift $\Delta F_0$ with color scale on the right) as a function of the length/coverage (horizontal axes) and temperature.  Quadratic fits $P=AT^2$ were used, with the pre-factor $A$ fitted as in panel (a). Red hyperbola is the onset of thermal excitation of phonons, $T_{ph}=\pi\hbar{c}/k_BL$. Note that thermal phonons contribute by order of magnitude less than zero-point phonons shown in Fig.\;\ref{fig:phonons0}}
\end{figure}

We demonstrate vanishing of phonons in our solid helium samples by measuring temperature dependence of pressure in the low temperature domain below the melting transition, see Fig.\;\ref{fig:phonons1}b. Indeed, the pressure is zero in small samples, and it appears only when the product $LT$ exceeds the onset condition given by $\pi\hbar{c}/k_B$. Note that, to reach a similar depletion of thermal phonons in a sample of 1\;mm size, a cool down to microKelvin temperatures would be required, which is not possible yet.

With increasing temperature (or, length), phonons start to enter the sample one by one, and thermodynamics should be revised, as number of phonons is so small. For instance, the pressure, which is the derivative of energy over the length, is essentially zero at most of the lengths except for special points $L=\pi\hbar{c}/k_BT, 2\pi\hbar{c}/k_BT, 3\pi\hbar{c}/k_BT$ etc., where the pressure becomes formally infinite. In small samples, obviously, energy can not be considered as a good thermodynamic potential, as its density is size-dependent. Moreover, the energy is not additive: when two samples join together, total energy becomes more than the sum of energies of the original samples, due to the appearance of additional long wave phonon modes. With further increase in length (temperature), thermodynamics reaches its classical form, $P=A{T^2}$.

Lacking better representation, we assume that the low temperature dependence $P(T)$ is quadratic, $P=AT^2$, with the pre-factor $A$ dependent on the length $L$. Accordingly, we have fitted temperature dependence of the central frequency as $\Delta{F_0}=A(L)T^2$, and the results of the fit are shown in the Fig.\;\ref{fig:phonons1}a. As one can see, the pre-factor $a$ is increasing by many orders of magnitude with the increase of length of the 1/3 phase. We interpret this as an expansion of the phase space of phonons.

\section{Size-dependent thermodynamics}

In Eqs.\;\ref{eq:thermo}, we have estimated thermodynamic quantities in the domain $Lk_BT\ll\pi\hbar{c}$ and shown that they are exponentially vanishing in this regime. Here we calculate exact expressions for the principal thermodynamic functions in the whole range of $L$ and $T$. We start from the statistical sum of a single phonon mode $k$, 
\begin{equation}
\label{eq:sum}
z_k=\sum_{v}e^{-E_v/(k_BT)}=\sum_{v}e^{-\hbar\omega_k(v+1/2)/(k_BT)}= \frac{e^{-\hbar\omega_k/(2k_BT)}}{1-e^{-\hbar\omega_k/(k_BT)}},
\end{equation}

\noindent
where $v$ is the occupation quantum number of $k$th mode \cite{LL5}. Free energy of a single mode is thus $f_k=-k_BT\ln{z_k}={\hbar\omega_k}/{2}+k_BT\ln{(1-e^{\hbar\omega_k/{k_BT}})}$, and the whole free energy of the sample supporting $m$ modes is 

\begin{eqnarray}
\label{eq:freeenergy}
F=\frac{m\hbar}{2}\sum_{k=\pi/L}^{\pi/a}\omega_k+mk_BT\sum_{k=\pi/L}^{\pi/a}\ln{(1-e^{\hbar\omega_k/k_BT})}= \nonumber \\
\frac{m\hbar{c}}{2}\sum_{k=\pi/L}^{\pi/a}k+\frac{mk_BTL}{\pi}\int_{\pi/L}^{\pi/a}\ln{(1-e^{\hbar{c}k/k_BT})}dk \cong \nonumber \\
\frac{m\pi\hbar{c}L}{4a^2}-\frac{mLk_B^2T^2}{2\pi\hbar c}\int_{\pi\hbar c/Lk_BT}^{\infty} \frac{tdt}{e^{t}-1},~~~~~~
\end{eqnarray}

\noindent
where $a$ is the interatomic distance that sets up a minimum possible wavelength of phonons $\lambda_{min}=2a$.

In macroscopic thermodynamics, no more than three phases can co-exist. It follows from the equalities of chemical potentials for the phases $\mu_1(P,T)=\mu_2(P,T)=\mu_3(P,T)$ which are two equations with two unknowns, $P$ and $T$. One cannot add another phase to the coexistence point because there will be three equalities and only two unknowns, which can be satisfied only accidentally. However, in the mesoscopic domain, $T\lesssim\pi\hbar{c}/L$, thermodynamics becomes more rich because here emerges an additional thermodynamic parameter, the length (volume) of the sample. Chemical potential, as all other thermodynamic potentials, is now a function of $L$, $\mu=\mu(P,T,L)$. As we add one more variable, three equalities can be satisfied simultaneously, $\mu_1(P,T,L)=\mu_2(P,T,L)=\mu_3(P,T,L)=\mu_4(P,T,L)$, and correspondingly four different phases may co-exist at a quadruple point. 

The differential of the chemical potential should now be written as $d\mu=vdP-sdT+v(\partial{P}/\partial{L})dL$ where the term $\partial{P}/\partial{L}$ is found as

\begin{eqnarray}
\label{eq:thermo1}
S=-\frac{\partial{F}}{\partial{T}}=  
\frac{mLk_B^2T}{\pi\hbar{c}}\int_u^{\infty} \frac{tdt}{e^t-1}+\frac{mk_B}{2}\frac{u}{e^u-1}, \nonumber \\
E=F+TS=\frac{\pi{m}\hbar{c}L}{4a^2}+\frac{mk_BT}{2}\frac{u}{e^u-1}-\frac{mLk_B^2T^2}{2\pi\hbar{c}} \int_u^{\infty}\frac{tdt}{e^t-1}, \nonumber \\
P=-\frac{\partial{F}}{\partial{L}}=  
\frac{mk_BT}{2L}\frac{u}{e^u-1}+\frac{mk_B^2T^2}{2\pi\hbar{c}}\int_u^{\infty} \frac{tdt}{e^t-1}, \\
\frac{\partial{P}}{\partial{L}}=  
\frac{m\pi\hbar{c}}{2L^3}~\frac{e^u(u-1)+1}{(e^u-1)^2} \nonumber
\end{eqnarray}

\noindent
where $u\equiv\pi\hbar{c}/(Lk_BT)$ is the universal thermodynamic factor reflecting mesoscopic effects. At $u\rightarrow0$ ($L\rightarrow\infty$ or $T\rightarrow\infty$), the temperature-dependent free energy in Eq.\;\ref{eq:freeenergy} simplifies to the standard macroscopic 1D expression $F=-(\pi Lk_B^2T^2)/(12\hbar c)$. However, the pressure $P$ acquires an additional term proportional to temperature, which is yet a small correction.
It is easy to show that $e^u(u-1)+1>0$ at any positive $u$, and therefore the derivative $\partial{P}/\partial{L}$ is always positive. Note that the pressure calculated in Eq. \ref{eq:thermo1} is not the Casimir pressure calculated in the first section but the pressure inside the sample, while the Casimir pressure is the difference between the pressure of the surrounding bath and that of the sample.

\begin{figure}[htb]
\includegraphics[width=0.98\linewidth]{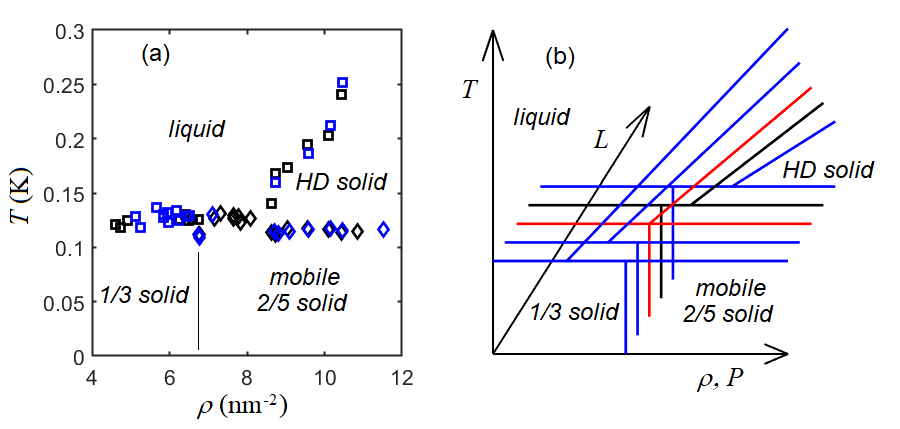}
\caption{\label{fig:quadrupol} Possibility of a quadruple point on the phase diagram. (a) A segment of the phase diagram of $^3$He on a nanotube \cite{CNT3HE} including liquid, 1/3 solid, 2/5 mobile solid, and high density (HD) solid. (b) Sketch of three-dimentional phase diagram. Increase of the sample length shifts the liquid--HD solid coexistence curve towards HD solid due to increasing pressure in the solid, see text for details. At a certain length the melting curve reaches the 1/3 -- 2/5 mobile solids coexistence curve forming a quadruple point (red diagram). Our measured phase diagram corresponds to the black diagram on the sketch. }
\end{figure}

Taking into account the additional thermodynamic parameter $L$, the phase diagram becomes three-dimensional as illustrated in the Fig.\;\ref{fig:quadrupol}(b). To be specific, we show phases observed in helium on nanotube: 1/3 solid, 2/5 mobile solid with delocalized vacancies, high density (HD) solid, and liquid \cite{CNT3HE}. With the increase of the length of the tube, the pressure in HD solid phase increases as discussed above. Therefore, the chemical potential of HD solid phase also increases, and the melting curve shifts to the right towards higher densities where pressure in the liquid is also larger. At certain length, the melting curve reaches the 1/3 -- 2/5 mobile solid coexistence line forming a quadruple point.

\section{Bose-Einstein condensation}
At large densities, $\rho>6.4$\;nm$^{-2}$, the 2/5 solid starts to form. This solid has been proposed to consist of bosonic dimers \cite{CNT3HE} and its lattice structure has topologically stable vacancies which arise owing to mismatch between the most symmetric helium structures and the carbon lattice. These vacancies exist even at zero-point and should Bose-condense at low enough temperatures \cite{AndreevLifshitz,Leggett1970,Chester1970,Boronat2012}. 

The transversal motion of an individual vacancy around the tube has a characteristic energy $\varepsilon_{\phi}=\hbar^2l^2/(2Mr^2)$ where $l$ is the angular momentum quantum number. The first excited state with $l=1$ and $M\sim2m_3$ has an energy of the order of 100\;mK, and therefore at lower temperatures the transversal motion is frozen, and the system of vacancies becomes {\it truly one-dimensional}. One can estimate the Bose-Einstein condensation temperature of a system of vacancies by adapting the standard approach for ideal bosonic gas (see, for instance Ref.~\onlinecite{LL3}) to the one-dimensional case. The number of particles will be given by a Bose-distribution integral with zero chemical potential,

\begin{eqnarray}
\label{eq:1D-1}
N=\frac{L}{2\pi\hbar}\int \, \frac{\mathrm{d}p}{\exp{\varepsilon/T}-1} 
=\frac{L\sqrt{MT}} {\sqrt{8}\pi\hbar} \int_{\varepsilon_{min}/T} ^\infty \frac{\mathrm{d}z}{\sqrt{z}(\exp{z}-1)},
\end{eqnarray}
 
\noindent
where $L$ is the length of the sample. Note that, in contrast to the 3D case, the integral diverges at low energies as $z^{-1/2}$, and the integration should start from the cutoff $\varepsilon_{min}/T=\hbar^2(2\pi/L)^2/(2M)$. The integral $I=\int_{\varepsilon_{min}/T} ^\infty z^{-1/2}\mathrm{d}z/(\exp{z}-1)$ in Eq. \ref{eq:1D-1} can be crudely estimated as follows: if $z\ll1$, it can be approximated as $\int_{\varepsilon_{min}/T} ^\infty z^{-3/2}\mathrm{d}z=(1/2)\sqrt{T/\varepsilon_{min}}$. However, when $z$ is of the order of 1 and larger, the integrand is significantly smaller than $z^{-3/2}$, and we may take $I\sim\sqrt{T/\varepsilon_{min}}=\sqrt{2MT}L/(2\pi\hbar)$, which yields

\begin{figure}[tbp]
\includegraphics[width=0.98\linewidth]{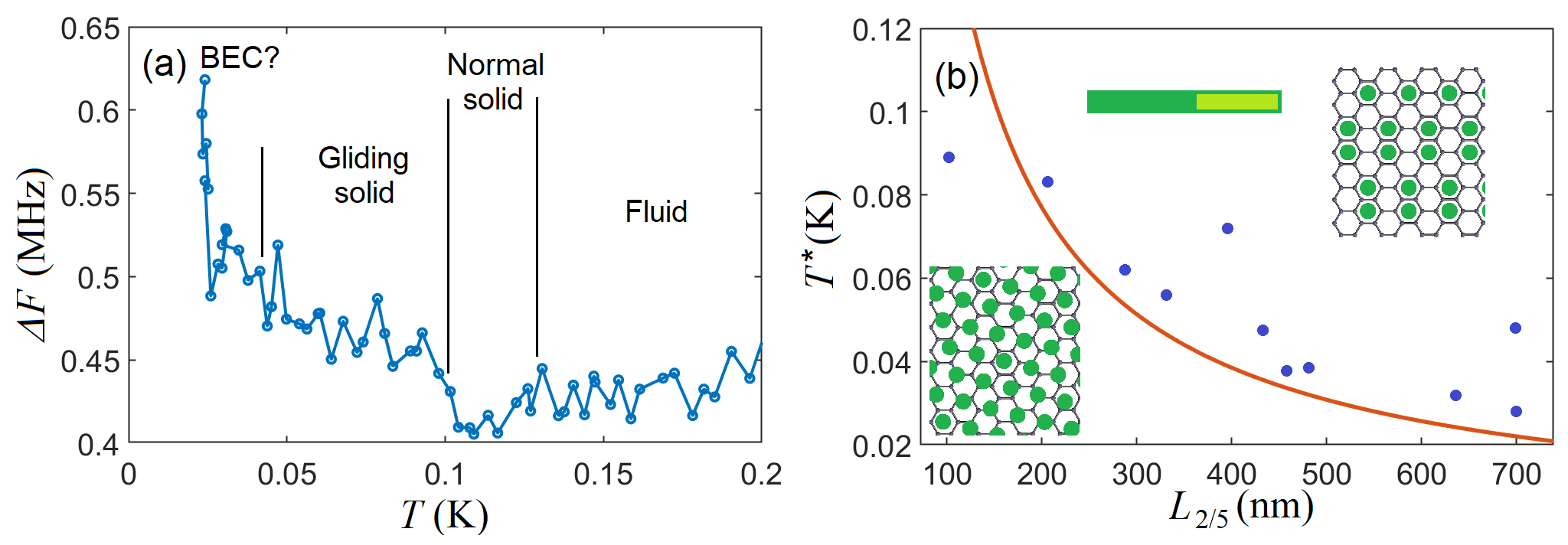}
\caption{\label{fig:dissipation_BEC} (a) Evidence for the Bose-Einstein condensation of vacancies. The dissipation increases substantially when temperature of mobile solid decreases below 50\;mK. Data taken with coverage 8.8\;nm$^{-2}$ corresponding to the length of the 2/5 mobile solid phase $L_{2/5}=L(\rho_{inc}-\rho)/(\rho_{inc}-\rho_{2/5})=460$\;nm (density of the incommensurate phase is $\rho_{inc}=11.0$\;nm$^{-2}$. (b) Possible onset temperature $T^*$ of Bose-condensation of delocalized vacancies as a function of the length $L_{2/5}$ occupied by the 2/5 solid phase, see text for details. Solid curve represents Eq.\;\ref{eq:1D-2} with mass $M=2m_3$ for the theoretical estimation of BEC transition. Inserts: sketches of the 2/5 phase (on the right) and of the incommensurate phase (on the left); horizontal bar illustrate partial occupation of the surface of the tube by the 2/5 and incommensurate solid.}
\end{figure}

\begin{equation}
\label{eq:1D-2}
T_{BEC}\sim\frac{4\pi^2\hbar^2n_{vac}}{ML},
\end{equation}

\noindent
where $n_{vac}=N/L$ is the linear density of vacancies. For the 2/5 solid shown in Fig.\;\ref{fig:dissipation_BEC}(b), $n_{vac}=1/(3a/2)=4.7$\;nm$^{-1}$. With the length of our system $L=700$\;nm, the condensation temperature would be in the range 10--20\;mK depending on the effective mass $M$ of the vacancies. We emphasize that Bose-Einstein condensation is possible in one-dimentional sample only due to the finiteness of the length $L$: in infinitely long 1D sample temperature of the condensation is zero.

At an arbitrary $^3$He coverage, a mixture of different phases is realized. However, two different commensurate phases cannot co-exist as they have different internal pressure. Indeed, density of commensurate phase cannot be changed at given temperature, therefore, pressure also is fixed, and mechanical equilibrium is not possible. At coverages 4.0\;nm$^{-2}<\rho<6.4$\;nm$^{-2}=\rho_{1/3}$ the commensurate 1/3 solid is coexisting with fluid. At higher coverages the 2/5 solid appears and coexists either with fluid or with incommensurate solid, depending on the filling factor. Clusters of different co-existing phases have smaller length than the length $L$ of the nanotube. In such reduced-size clusters of gliding solid, delocalized vacancies will Bose-condense even at higher temperatures, according to Eq.\;\ref{eq:1D-2}, forming a supersolid. 

The expected BEC transition would manifest itself by even stronger dissipation in the gliding solid phase at the lowest temperatures. Indeed, most of the measured $\Delta{F}(T)$ traces show significant increase of the dissipation $\Delta{F}$ when lowering temperature of the gliding solid (see Fig.\;\ref{fig:dissipation_BEC}).

The length of the 2/5 "gliding"solid phase in contact with the incommensurate phase can be calculated using the fact that pressure remains constant, as density of commensurate phase cannot be changed with filling. Therefore, density of incommensurate phase also doesn't change with filling and equals to maximum first layer coverage, $\rho_{inc}=\rho_{max}\simeq11.0$\;nm$^{-2}$ \cite{Lauter1990a,CNT3HE}. With the known density of both phases it is easy to calculate the lengths of both phases for a particular filling. For example, 1:3 mixture of 2/5 "gliding" solid and incommensurate solid is realized at the coverage $\rho=\rho_{2/5}\rho_{inc}/[(x(\rho_{inc}-\rho_{2/5})+\rho_{2/5}]=9.9$\;nm$^{-2}$ ($\rho_{2/5}=7.64$\;nm$^{-2}$, $x=0.25$). The cluster of 2/5 solid will have a length $L_{2/5}=x\rho_{inc}L/[(x(\rho_{inc}-\rho_{2/5})+\rho_{2/5}]=240$\;nm, and the Bose-Einstein condensation temperature of 45\;mK.

Due to a lag of helium motion behind the oscillating nanotube, friction occurs resulting in an additional dissipation. This additional dissipation is proportional to the mobility $\mu$ of helium, similarly to an electrical circuit in which the current $I$ and the power dissipation $P$ are proportional to the conductance $G$, $I=GU$, $P=GU^2$. The high dissipation in the low temperature phase at elevated coverages is one of the strongest arguments for the model of "gliding" solid \cite{CNT3HE}. However, when mobility increases further, to infinity in the supersolid case, dissipation should drop again because the relaxation time will become much longer than the oscillation period. Indeed, the oscillating tube drags helium along by the drag force
$f_d=(\dot{x}_{CNT}-\dot{x}_{He})/\mu=m_{He}\ddot{x}_{He}$ where $x_{CNT}=X\exp{i\omega{t}}$ and $x_{He}=x\exp{i\omega{t}}$, and $m_{He}$ is effective mass of helium. Amplitude of oscillations of adsorbed helium turns out to be always smaller than the amplitude of the tube, $x=X/(1+i\omega{m}_{He}\mu)$.
Average dissipation due to the drag is found as 

\begin{eqnarray}
\overline{P}=\overline{f_d\cdot(\dot{x}-\dot{X}}) =m\overline{\ddot{x}(\dot{x}-\dot{X}}) 
=\frac{1}{2}\frac{m_{He}^2X^2\omega^4\mu}{1+(\omega{m}_{He}\mu)^2} \nonumber.
\end{eqnarray}

Energy of the oscillating nanotube is ${E}={E_{kin}+E_{pot}}=2\overline{E_{kin}}=(1/2)MX^2\omega^2$, and the dissipation rate is therefore 

\begin{equation}
\label{eq:tau}
\Gamma=2\pi\Delta{F}=\overline{P}/{E}=(m_{He}/M)\cdot\omega^2\tau/(1+\omega^2\tau^2)
\end{equation}

\noindent
where $\tau=m_{He}\mu$ is the relaxation time of helium on nanotube. The case of short $\tau$ corresponds to a solid phase in which helium is frozen to the carbon lattice. Dissipation due to mutual motion of helium and nanotube is minimal as there is no or very small mutual friction. When helium melts to a fluid, the relaxation time increases together with the dissipation rate. More interesting is the case of mobile zero-point vacancies in the range $\rho>6.4$\;nm$^{-2}$ where three regimes are possible. At temperatures above 0.1\;K vacancies are localized \cite{CNT3HE}, and solid helium behaves as normal stiff crystal which oscillates in phase with the tube and has therefore low dissipation ($\tau\rightarrow0$), see Fig.\;\ref{fig:dissipation_BEC}. When temperature is lowered below 0.1\;K, vacancies becomes delocalized providing a finite mobility and relaxation time of solid. This so-called "gliding solid" is lagging behind the nanotube, which provides even higher  dissipation $\Gamma$ than in the liquid. On further cooling, dissipation increases even more which we attribute to partial Bose-Einstein condensation of vacancies. Indeed, vacancies condensed into the ground state are superfluid, and effective mobility increases. However, at $T\rightarrow0$ where the condensation of vacancies is nearly complete, mobility and relaxation time go to infinity, and, according to Eq.\;\ref{eq:tau}, the dissipation also goes to zero; in this regime vacancies are not moving at all. 

From the above analysis it follows that there must be a maximum of dissipation rate $\Gamma$ at $\omega\tau=\omega{m}\mu=1$ where helium and the nanotube oscillate in counter-phase. Therefore, the increase of dissipation at $T<40$\;mK seen in Fig.\;\ref{fig:dissipation_BEC} should change to a decrease when significant part of vacancies will be in the ground state. Observation of such a bump in temperature dependence of the dissipation is an evidence of the Bose-Einstein condensation of vacancies.

\begin{figure}[htb]
\label{fig:01mar}
\includegraphics[width=0.98\linewidth]{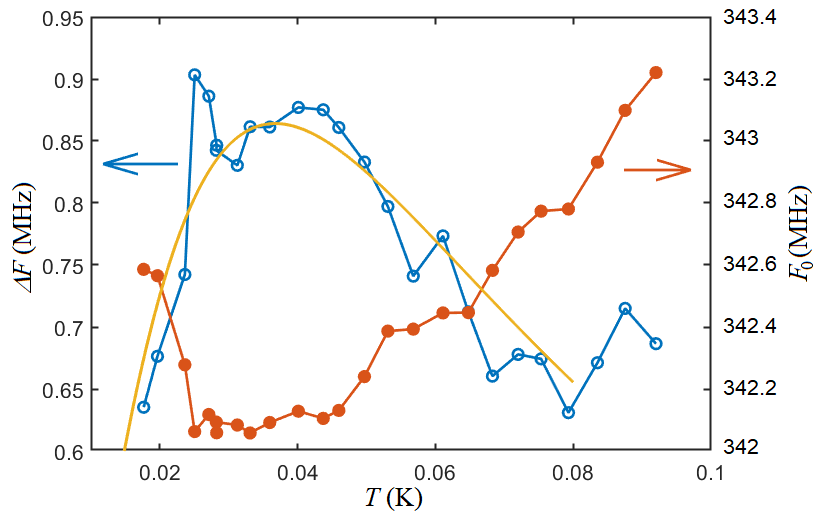}
\caption{Dissipation (open circles) and central frequency (filled circles) of the resonance with the sample of the length $L_{2/5}=210$\;nm ($\rho_{2/5}=6.74$\;nm$^{-2}$). The curve is of fit of dissipation-induced $\Delta{F}$ using Eq.\;\ref{eq:BEC_fit} and $m_{vac}/M=2(\Delta{F}_{max}-\Delta{F}_b)/F_0=5.1\cdot10^{-4}$ (see text). Increase of the central frequency at lowest temperatures by 0.55\;MHz corresponds to the relative mass decoupling $m_{He}/M=3.2\cdot10^{-3}$.}
\end{figure}

Generally, in 1D-case the number of vacancies in the excited state is proportional to temperature, $n_{ex}/n_{vac}=T/T_{BEC}$. In the normal state, $T\geq{T_{BEC}}$, the relaxation time $\tau$ is finite and determined by the friction coefficient $m_0/\tau_0$ of individual vacancies: $m_{vac}/\tau_{norm}=N_{vac}m_0/\tau_0$. Below $T_{BEC}$, vacancies in the condensate don't contribute to the friction, and friction coefficient decreases correspondingly 
$1/\tau=N_{ex}/\tau_0=(N_{ex}/N_{vac})N_{vac}/\tau_0=(T/T_{BEC})/\tau_{norm}$. After substituting this scaling relation to Eq.\;\ref{eq:tau}, we obtain the desired expression for the width of the resonance below $T_{BEC}$:

\begin{equation}
\label{eq:BEC_fit}
2\pi[\Delta{F}(T)-\Delta{F}_{b}]=
\omega\frac{m_{vac}}{M}\frac{\omega\tau_{norm}(T_{BEC}/T)}
{1+\omega^2\tau_{norm}^2(T_{BEC}/T)^2},
\end{equation}

\noindent
where $\Delta{F}_b$ is the width of the resonance of the bare nanotube.  In the Fig.\;\ref{fig:01mar} we show the temperature dependence of the linewidth at density 6.74\;nm$^{-2}$ ($L_{2/5}=206$\;nm). The fit of the experimental data with the Eq.\;\ref{eq:BEC_fit} is excellent and gives reasonable values of the relaxation time, $\tau=0.22$\;ns, and of the relative mass of vacancies $m_{vac}/M=5.1\cdot10^{-4}$.

Deep in the BEC-state, vacancies decouple solid helium from the tube, and the effective mass of the resonator decreases with a corresponding increase in the central frequency, $\Delta F_0=(F_0/2)(m_{He}/M)$. The shift of the resonance at $\rho=6.74$\;nm$^{-2}$ is about 0.55\;MHz which corresponds to relative decoupled mass $m_{He}/M=3.2\cdot10^{-3}$. As 2/5 solid consists of dimers, we assume that the mass of single vacancy is twice the mass of helium atom which gives the concentration of vacancies in the dimer 2/5 solid as 
$n_{vac}=(m_{vac}/M)/(m_{He}/M)=16$\;\%. Assuming one vacancy per period of the 2/5 solid along the tube, and 4-5 periods perpendicular to the axis of the tube (see insert to Fig.\;\ref{fig:dissipation_BEC}b), we have one vacancy per 8-10 dimer sites which gives the concentration of zero-point vacancies about 10--15\;\%, perfectly matching the value obtained from the fits \footnote{The helium lattice period along the tube is two rows, and, assuming one vacancy per elementary cell, we have one vacancy per 8-10 dimer sites which gives the concentration of zero-point vacancies about 10--15\;\%. The mass of decoupled helium is also reliable, as the length of the sample is about 30\;\% of the total length of the tube, and for 2/5 solid the number of helium atoms is 4/10 of number of carbon atoms (we ignore number of vacancies) and taking into account atomic masses we find $m_{2/5}/M=1.3\cdot10^{-2}$. This value differs by only four times from the value estimated from line shift, and this difference can be explained by the non-symmetric position of the 2/5 phase on the tube.}.

The BEC of vacancies and the acoustic Casimir effect have very much in common. Indeed, both particles obey Bose-statistics and have certain minimum thermal energy due to the restricted geometry. Both systems therefore become degenerate, when the length of the sample decreases. In the case of vacancies, Bose-Einstein condensation temperature is inversely proportional to the length $L$, see Eq.\;\ref{eq:1D-2}. On their part, phonons condense in the ground state at temperature $T_{C}\approx\pi\hbar{c}/L$ which differs, besides numerical factors, by the ratio $c/(\hbar{n_{vac}}/M)$. Note that, the combination $\hbar{n_{vac}}/M$ has a meaning of average zero-point velocity of vacancy. Indeed, according to Heisenberg uncertainty principle, characteristic momentum of the particle is $p\sim\hbar/\delta{x}$ where uncertainty of the coordinate is of the order is an inverse linear density, $\delta{x}\sim1/n_{vac}$. After dividing by mass of vacancy $M$ we end up with the velocity of vacancy. Thus, the degeneracy in both systems, phonons and vacancies, has inherently the same origin: it happens when temperature decreases below lowest possible frequency in the system, $T<\hbar\omega_{min}\sim\hbar({v}/L)$.

Indeed quite analogously, there exists zero-point energy of vacancies and zero-point energy of phonons in our experimental object. The statistics of the two systems is identical, and the number of (quasi-)particles is conserved in both systems: number of vacancies is set by the topological invariant of the mismatch between the carbon lattice and helium superlattice \cite{CNT3HE}, while the number of phonon modes is set by the number of helium atoms and the geometry of the tube. 

Helium on nanotube is therefore a unique setting, which is small enough to manifest mesoscopic thermodynamics already at temperatures below 100\;mK which are available in a regular dilution cryostat. It would be extremely interesting to cool the $^3$He-on-nanotube system further towards lower temperatures to investigate the fully condensed state of zero-point vacancies. Intriguingly, the second layer of helium on CNT promises various new phases as it is very weakly coupled to the carbon lattice. The proposed length-dependent quadruple point could be demonstrated for the first time using $^3$He on nanotube. Furthermore, subatomic layers of $^4$He on carbon nanotube could prove very interesting objects to study as zero-point vacancies in this case are lighter than in the dimer $^3$He solid, so that their Bose-condensation is expected to take place at even higher temperatures.

\section*{Appendix: electromechanical scheme and measurements of coverage.}

When RF voltage $\Delta U=U_{RF}\cos{\Omega t}$ is applied to the tube, it starts to oscillate because of the oscillating electrical field $\Delta E=\Delta U/H$, where $H$ denotes the distance between back gate and the CNT. With gate-induced charge $Q$, the oscillating electrical force $F=Q\cdot\Delta E$ will produce mechanical motion $h=H-H_\mathrm{eq}$ of the tube from the equilibrium position $H_\mathrm{eq}$, governed by the equation of motion $CU_{gate}U_{RF}\cos{\Omega t}/(r_0\log{2H/r_0})=M\partial^2h/\partial t^2+\gamma\partial h/\partial t+kh$, where $C=2\pi\varepsilon_0L/\log{(2H/r)}$ is the capacitance of the suspended tube, $U_{gate}$ is the DC gate voltage, and $M$ is the effective mass of the tube. The conductance $G(E)$ of the tube also oscillates at the same frequency, as it is field-dependent. As a result, the current through the tube $I=\Delta U G \propto\cos^2{\Omega t}$ has a constant term which is proportional to the amplitude of the mechanical oscillations \cite{Gouttenoire2010}. If the RF frequency is modulated by a much lower frequency $\omega\sim 1$\;kHz, $\Omega=\Omega+\delta\cos{\omega t}$, the current will have a component at low frequency,

\begin{equation}
\label{eq:fit}
<I>_{\omega}\propto\frac{d}{d\Omega}~\frac{1}{\Omega_0^2-\Omega^2+2\pi i{\it\Gamma}\Omega},
\end{equation}

\noindent
which can be detected using a low-frequency lock-in amplifier. Here $\Omega_0=2\pi{F_0}$ is the central circular frequency and $\Gamma=2\pi\Delta{F}$ is the dissipation. In the method described in Ref.\;\onlinecite{Gouttenoire2010}, only the real part of the differentiated Lorentzian in Eq. (\ref{eq:fit}) is considered. In the more general case, due to finite capacitance of the tube and the electrodes, there is a phase shift between the voltage and conductance, and the uncertainty in the phase of the Lorentzian response needs to be taken into account.

\section*{Acknoledgements}

We are grateful to Vladimir Eltsov for fruitful discussions. This work was supported by the Academy of Finland (AF) projects 341913 (EFT). The research leading to these results has received funding from the European Union’s Horizon 2020 Research and Innovation Programme, under Grant Agreement No.~824109 (EMP). The experimental work benefited from the Aalto University OtaNano/LTL infrastructure. We are grateful to Petri Tonteri from Densiq Ltd.~(90620 Oulu, Finland) for providing us with the ultra-pure grafoil \cite{grafoil}.

\end{document}